\documentclass[10pt,journal,compsoc]{IEEEtran}

\usepackage{subcaption}
\usepackage[table]{xcolor}
\usepackage{algorithm,algorithmic}
\usepackage{eqparbox}

\usepackage{tikz}
\makeatletter
\let\@tabclassz =\@classz
\let\@tabclassiv =\@classiv
\makeatother
\definecolor{lightgray}{gray}{0.9}
\definecolor{lightblue}{rgb}{0.96,0.96,1.0}

\usepackage[cmex10]{amsmath}
\usepackage{nicefrac}
\usepackage{ragged2e} 
\usepackage[normalem]{ulem}
\usepackage{wrapfig}
\usepackage{amssymb}
\usepackage{bm}
\usepackage{listings}
\usepackage{multirow}

\usepackage{graphicx}
\graphicspath{{images/}}

\usepackage{color}

\newcommand{\revise}[2]{#2}

\usepackage{hyperref}
\usepackage{xcolor}
\hypersetup{
    colorlinks,
    linkcolor={red!50!black},
    citecolor={blue!50!black},
    urlcolor={blue!80!black}
}

\begin{document}

\title{Infill Optimization for Additive Manufacturing \\ --Approaching Bone-like Porous Structures}

\author{Jun Wu, Niels Aage, R\"udiger Westermann, Ole Sigmund
\IEEEcompsocitemizethanks{
\IEEEcompsocthanksitem
Jun Wu, Niels Aage and Ole Sigmund
are with the Department of Mechanical Engineering,
Technical University of Denmark, Lyngby, Denmark.\protect\\
\IEEEcompsocthanksitem R\"udiger Westermann is with the Department of Computer Science, Technische Universit\"at M\"unchen, Munich, Germany. \protect\\
\IEEEcompsocthanksitem
Corresponding Author: Jun Wu, E-mail: junwu@mek.dtu.dk
}
\thanks{}}

\markboth{Wu \MakeLowercase{\textit{et al.}}: Infill Optimization for Additive Manufacturing}
{Wu \MakeLowercase{\textit{et al.}}: Infill Optimization for Additive Manufacturing}

\IEEEtitleabstractindextext{%
\begin{justify}
\begin{abstract}
Porous structures such as trabecular bone are widely seen in nature. These structures exhibit superior mechanical properties whilst being lightweight. In this paper, we present a method to generate bone-like porous structures as lightweight infill for additive manufacturing. Our method builds upon and extends voxel-wise topology optimization. In particular, for the purpose of generating sparse yet stable structures distributed in the interior of a given shape, we propose upper bounds on the \textit{localized} material volume in the proximity of each voxel in the design domain. We then aggregate the local per-voxel constraints by their \textit{p}-norm into an equivalent global constraint, in order to facilitate an efficient optimization process. Implemented on a high-resolution topology optimization framework, our results demonstrate mechanically optimized, detailed porous structures which mimic those found in nature. We further show variants of the optimized structures subject to different design specifications, and analyze the optimality and robustness of the obtained structures.
\end{abstract}
\end{justify}

}

\maketitle

\IEEEdisplaynontitleabstractindextext

%
\IEEEpeerreviewmaketitle

\section{Introduction}
The term \textit{infill} in additive manufacturing (also known as 3D printing) refers to the interior structure of an object that is printed. It often has a \textit{regular} pattern, which is selected by the user in the slicing software, along with a specific volume percentage. The infill pattern and percentage significantly influence the printing process, as well as physical properties of the printed object. In general, a higher volume percentage leads to a print that is more resistant to external loads, while consuming more material and prolonging the print time. To assist users in designing lightweight but mechanically strong prints, it is highly interesting to resort to structural analysis and optimization to find an optimal layout of the interior structure, which goes beyond the regular patterns.

\begin{figure*}[t]
\centering
\includegraphics[width=0.22\linewidth]{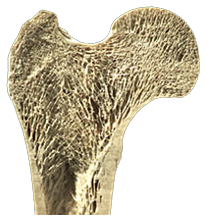}
\includegraphics[width=0.22\linewidth]{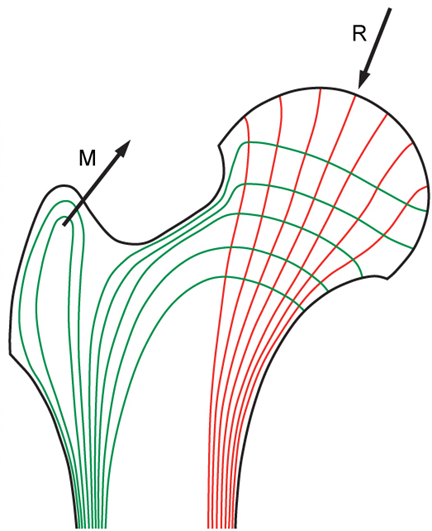}
\includegraphics[trim = 0mm 40mm 0mm 0mm, clip, width=0.25\linewidth]{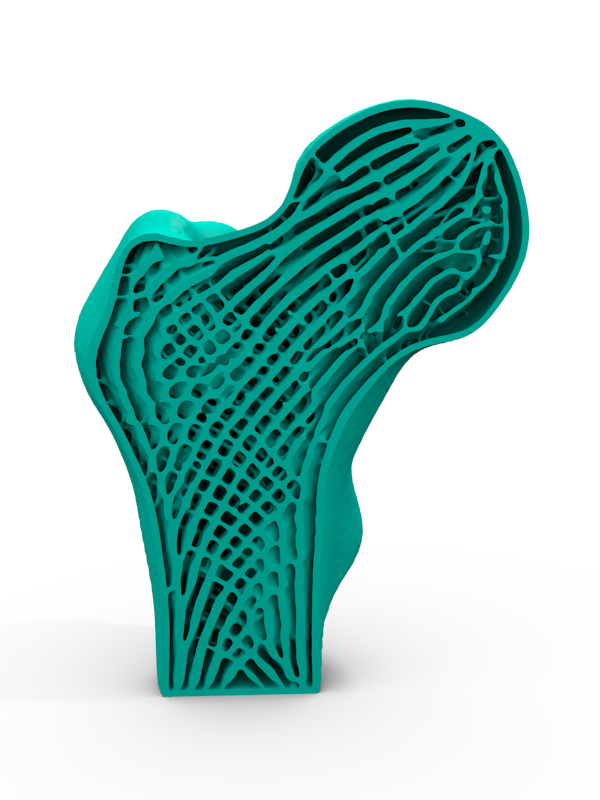}
\includegraphics[trim = 0mm 10mm 0mm 0mm, clip, width=0.22\linewidth]{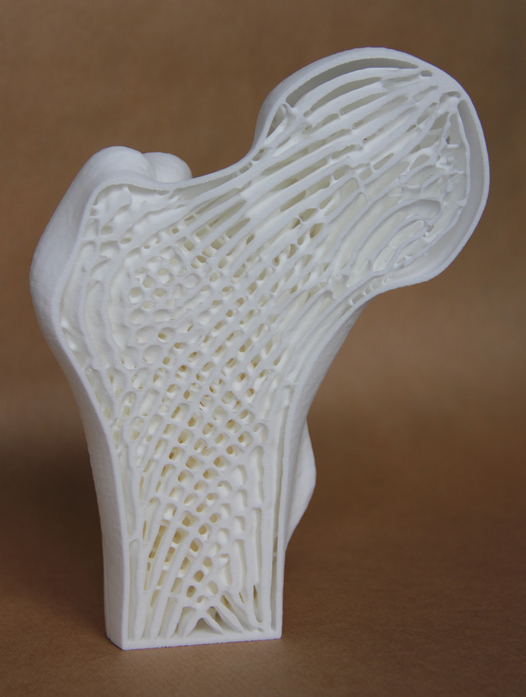}
\caption{From left to right: Cross-section of a human femur showing cortical structures on the shell and trabecular structures in the interior~\cite{Bone}. Illustration of principle stress directions under major mechanical loads~\cite{Dick09}. Cross-section of the optimized porous infill in a 3D bone model. The 3D printed bone model.
}\label{fig:femur}
\end{figure*}

Our research regarding optimal infill is inspired by the architecture of bone. Bone is \revise{composited}{composed} of two types of structures -- compact \textit{cortical bone} forming its outer shell, and spongy \textit{trabecular bone} occupying its interior (see the cross section of a human femur in Fig.~\ref{fig:femur}). This composite results from a natural optimization process, during which the bone adapts itself to the mechanical load (Wolff's law~\cite{Wolff1893}). As a consequence of this adaptation, micro-structures of trabecular bone are aligned along the principle stress directions as illustrated in the second image of Fig.~\ref{fig:femur}. This natural optimized composition is lightweight, resistant, robust with respect to force variations, and damage-tolerant~\cite{Meyers08,Wegst15}. These properties make bone-like structure an appealing option as infill for additive manufacturing.

In this paper we present an approach for the generation of bone-like porous structures. Our approach builds upon and extends the general, voxel-wise topology optimization scheme~\cite{Bendsoe88,Bendsoe04}. It maximizes the mechanical stiffness by optimizing the distribution of a prescribed amount of material in a given design domain, under a given set of external loads. In particular, to generate porous structures, we propose a formulation to measure local volume fractions, and then impose constraints on such values in order to regulate the local material distribution. Additionally, under the objective function to maximize stiffness, the porous structures are automatically aligned to accommodate the mechanical loads in an optimized manner.

The specific contributions of our paper include:
\begin{itemize}
\item A novel formulation for generating porous structures firmly based on structural optimization, and
\item Insights into optimal structures from a mechanical perspective, verified by a detailed parameter study. 
\end{itemize}

The remainder of this paper is organized as follows. After reviewing related work in Section~\ref{sec:relatedWork}, we present in Section~\ref{sec:method} the problem formulation and the techniques for solving the infill optimization problem. In Section~\ref{sec:extension}, we discuss extensions to steer the optimization process. Results and analysis are presented in Section~\ref{sec:results}, before conclusions are drawn in Section~\ref{sec:conclusion}.

\section{Related Work}
\label{sec:relatedWork}

With the ever-increasing popularity of consumer 3D printers, much research has been devoted to address geometric and physical modeling problems for computational fabrication, \revise{}{including the toolpath generation~\cite{Zhao16}}. In this section, we review techniques related to the optimization of mechanical properties. For an overview of geometric and physical modeling for 3D printing, let us refer to a recent survey article~\cite{Gao15Survey} and a tutorial~\cite{Liu14}. 

To assist users in the design of 3D printed shapes, \revise{Stave}{Stava} et al. ~\cite{Stava12} presented a system to detect structural deficiencies by finite element analysis. A set of correction operations including hollowing, thickening, and strut insertion, are proposed to improve the structural soundness. Targeting specifically for reducing the material usage inside a 3D model, Wang et al.~\cite{Wang13} introduced skin-frame structures as a composition of the shape interior, and optimized the layout and size of frames. Lu et al~\cite{Lu14} proposed  honeycomb-like Voronoi structures to hollow the interior volume, and optimized the distribution and size of Voronoi cells. In early work by Smith et al.~\cite{Smith:2002:CMT:566654.566580}, the layout of truss structures is optimized for designing bridges and towers. \revise{}{Manufacturing constraints regarding the avoidance of overhang surfaces and small geometric feature size, have been addressed by Wu et al.~\cite{Wu16CAD} via self-supporting rhombic structures for infill optimization. Manufacturable micro-structures have been investigated in graphics~\cite{Schumacher15,Panetta15,Martinez16} and in mechanical topology optimization approaches (e.g., ~\cite{Andreassen14,Alexandersen15}, among others).} Our work is inspired by these works\revise{,}{} \revise{but}{and} builds upon topology optimization which doesn't prescribe the structural composition \textit{a priori} and thus does not limit the design space.

\noindent \textbf{Topology optimization} \quad
Topology optimization is based on a volumetric element-wise parametrization of the design domain. This general formulation does not prescribe the topology \textit{a priori}, but allows structures to appear and adapt during the iterative optimization process. For a thorough review of topology optimization techniques, let us refer to recent survey articles~\cite{Sigmund13,Deaton14}. Our work is based on the density approach, which is known as Solid Isotropic Material with Penalization (SIMP)~\cite{Sigmund01}. The method is related to the length scale problem in the literature of topology optimization, where the interest is to control the minimal and/or maximal structure size for manufacturability~\cite{Poulsen03,Lazarov16}. In particular, we follow the idea of the projection filter~\cite{Guest04,Guest09} in our implementation to impose local volume constraints. Different from exact length scale control, we propose a projection method in an approximate manner which facilitates fast numerical solution. In our work we employ the potential of numerical multigrid schemes to enable topology optimization at high resolution and efficiency~\cite{Wu16}.

\noindent \textbf{Approaching bone-like structures} \quad
While we approach bone-like porous structures for their superior mechanical properties by using topology optimization, we note that other directions---considering different aspects of bone---exist for generating such structures. One such direction is material reconstruction. For instance, Liu and Shapiro~\cite{Liu15} proposed to reconstruct 3D micro-structures from 2D sample images using example-based texture synthesis~\cite{Wei09}, such that the synthesised structures preserve statistical features of the given sample. 

Another direction is the simulation of bone tissue adaptation using a biological model. For instance, Huiskes et al.~\cite{Huiskes00} proposed a biological model to simulate the process of bone resorption and formation under given mechanical stimuli, following Wolff's bone re-modeling theory~\cite{Wolff1893}. Numerical modeling of bone adaptation has been further studied in computational mechanics by applying two-scale simulations (e.g.,~\cite{Coelho09,Wang16,Schury12}), which date back to the seminal work by Bends{\o}e and Kikuchi~\cite{Bendsoe88}. The two-scale approach combines a fine scale with predefined optimal, single or multi-scale micro-structures which are mechanically characterized by numerical or analytical homogenization, with a coarse scale guided by finite element analysis and topology optimization. Besides the challenge of obtaining continuous micro-structural details between neighbouring cells, the generated structure is typically a regular repetition of (a limited amount of types of) cells. In contrast to such approaches which strictly separate local and global scales, we propose to control local details by embedding geometric constraints into a unified simulation and optimization scale\revise{}{, similar to the approach by Alexandersen et al.~\cite{Alexandersen15}. In contrast, however, our approach provides higher geometrical flexibility}. As a result, the resultant structure varies smoothly across the entire domain, and its geometric features are not restricted to a set of prescribed cell types.

\section{Infill Optimization}
\label{sec:method}

We start by formulating a discrete optimization problem to generate porous structures, then introduce relaxations for numerically solving the problem, followed by a summary of the algorithm. Following the techniques, we present a 2D example to demonstrate and explain the consequences of this formulation for the resulting structures.

\subsection{Discrete Formulation}
Our formulation is based on a regular hexahedral discretization of the design domain $\Omega$ that is covered by a given shape. For each volumetric element (i.e., voxel) $e$ in the discretized model, a boolean value $\rho_e \in\{0,1\}$ is assigned to indicate a solid voxel ($\rho_e = 1$), or an empty one ($\rho_e=0$). This leads to a binary field $\rho$ representing the material distribution in $\Omega$. 

We define $\overline{\rho}_e$ to quantify the (local) material distribution in a neighbourhood surrounding the voxel $e$. In particular, $\overline{\rho}_e$ measures the percentage of solid voxels over all voxels in a prescribed neighbourhood $\mathbb{N}_e$, i.e.,
\begin{equation}
\overline{\rho}_e = \frac{\sum_{i\in \mathbb{N}_e} \rho_i}{\sum_{i\in \mathbb{N}_e} 1}.
\label{eq:localDensity}
\end{equation}
$\mathbb{N}_e$ is the set of all surrounding voxels with a centroid that is closer than a given influence radius $R_e$ to the centroid of voxel $e$, i.e., 
\begin{equation}
\mathbb{N}_e=\{i| \; ||x_i-x_e||_2 \le R_e\},
\label{eq:localMeasure}
\end{equation}
where $x_i$ and $x_e$ are the centroids of the voxels. The positions and lengths are measured in the unit of voxel. A local volume percentage $\overline{\rho}_e = 0.0$ (resp. $\overline{\rho}_e = 1.0$) means that all voxels in the defined neighbourhood are empty (resp. solid), and a value between $0.0$ and $1.0$ means that both empty and solid voxels exist.

With $\rho$ and $\overline{\rho}$ defined, the optimization problem is given as
\begin{align}
\underset{\rho}{\text{min}}
&\quad c = \frac{1}{2} u^T K u, \label{eq:objective} \\
\text{s.t.}
&\quad Ku = f,  \label{eq:state} \\
&\quad \rho_e \in \{0,1\},  \; \forall e, \label{eq:design} \\
&\quad \overline{\rho}_e \le \alpha, \; \forall e. \label{eq:localConstraint}
\end{align}
The objective is to minimize the compliance, measured by the strain energy $c$, with $u$ being the displacement vector, and $K$ being the stiffness matrix. The displacement vector $u$ is obtained by solving the static elasticity equation Eq.~\ref{eq:state} under the external force vector $f$. Eq.~\ref{eq:design} restricts the design variables to take discrete values $0$ (empty) or $1$ (solid).

The novel part in our formulation is Eq.~\ref{eq:localConstraint}. This constraint restricts the local material accumulation. For instance, $\alpha = 0.6$ means at most $60\%$ of voxels in $\mathbb{N}_e$ are solid, while the other $40\%$ are empty. Note that while this constraint restricts the percentage of the solid/empty voxels, it does not prescribe which specific voxels are solid or empty: Determining the specific solid and empty voxels is left to the optimizer, under the goal of reducing the objective function. The rational behind this constraint is that it prevents material from being accumulated to form large solid regions, and as a consequence, the material will be distributed more evenly over the domain. This is in line with what we observe in nature when looking at porous structures such as trabecular bone.

We note that additional constraints such as a maximum total volume known from classical topology optimization can be integrated into this formulation as well. Its influence on the resulting structures, as well as the influence of other parameters will be discussed in Section~\ref{sec:extension}.

\subsection{Relaxations}

The optimization problem given in Eq.~\ref{eq:objective}-\ref{eq:localConstraint} is a discrete optimization problem, with up to millions of variables (cf. per-voxel design variable in Eq.~\ref{eq:design}) and up to millions of constraints (cf. per-voxel constraint in Eq.~\ref{eq:localConstraint}) in some of the test models. In the following, we present relaxations to approximate this problem and facilitate numerical optimization.

\subsubsection{Constraint Aggregation}
The per-voxel local volume constraint (Eq.~\ref{eq:localConstraint}) gives rise to a large number of constraints. These constraints are equivalent to $\max\limits_{\forall e}(\overline{\rho}_e) \le \alpha$, which reduces the large number of constraints into a single constraint. However, it is not differentiable, and thus not directly applicable to numerical optimization schemes. To overcome this problem, we use the $p$-norm function to approximate the $\max$ function,
\begin{equation}
\max\limits_{\forall e}(\overline{\rho}_e) \approx ||\overline{\rho}||_p = (\textstyle \sum\limits_{e} \overline{\rho}^p_e)^{\frac{1}{p}}.
\end{equation}
As $p$ goes to infinite, $||\overline{\rho}||_p$ becomes equivalent to $\max\limits_{\forall e}(\rho_e)$. To account for the difference between $\max\limits_{\forall e}(\overline{\rho}_e)$ and $||\overline{\rho}||_p$ when the value of $p$ is not infinitely large, we write the consolidated constraint $\max\limits_{\forall e}(\overline{\rho}_e) \le \alpha$ by 
\begin{equation}
(\textstyle \sum\limits_{e} \overline{\rho}^p_e)^{\frac{1}{p}} \le  (\textstyle \sum\limits_{e} \alpha^p)^{\frac{1}{p}},
\end{equation}
which can be rearranged to
\begin{equation}
\textstyle{ ( \frac{1}{n} \sum\limits_{e} \overline{\rho}^p_e ) ^{\frac{1}{p}} } \le \alpha,
\end{equation}
where $n$ is the number of elements. A larger $p$ more strictly enforces the per-voxel constraints, while increasing the non-linearity of the problem. In our examples we choose $p=16$. 

\subsubsection{Continuous Design Variable, Filtering, and Projection}
The discrete design variable (Eq.~\ref{eq:design}) necessitates expensive integer programming. To facilitate efficient gradient-based numerical optimization, we follow the study in~\cite{Wang10} and introduce a per-voxel design variable $\phi_e$ which is allowed to take a scalar value continuously varying between $0.0$ and $1.0$,
\begin{equation}
\phi_e \in [0.0, \; 1.0].
\end{equation}
The field of design variables $\phi$ is first smoothed via a convolution filter. The filtered field $\tilde{\phi}$ is then projected to get the material distribution $\rho$. These two steps involve only local operations, and are called filtering and projection, respectively.

\noindent \textbf{Filtering $\phi \to \tilde{\phi}$} \quad 
The purpose of the filtering $\phi \to \tilde{\phi}$ is to remove checkerboard patterns (i.e., regions of alternating solid and void voxels) resulting from numerical instabilities~\cite{Diaz95}. In particular, the local filter calculates a weighted average of the neighbouring values,
\begin{equation}
\tilde{\phi}_e = \frac{\sum_{i\in\mathbb{M}_e} \omega_{i,e} \phi_i}{\sum_{i\in\mathbb{M}_e} \omega_{i,e}},
\label{eq:densityFilter}
\end{equation}
where $\mathbb{M}_e$ is the set of voxels close to voxel $e$, defined by
\begin{equation}
\mathbb{M}_e = \{i| \; ||x_i-x_e||_2 \le r_e\}.
\end{equation}
Here, $r_e$ defines the filter radius. This filter size is different and smaller than the radius $R_e$ in Eq.~\ref{eq:localMeasure}. The weighting factor $\omega_{i,e}$ linearly depends on the distance between the considered voxels,
\begin{equation}
\omega_{i,e} = 1 - \frac{||x_i-x_e||_2}{r_e}.
\end{equation}

\begin{figure}[h]
\centering
\includegraphics[width=0.7\linewidth]{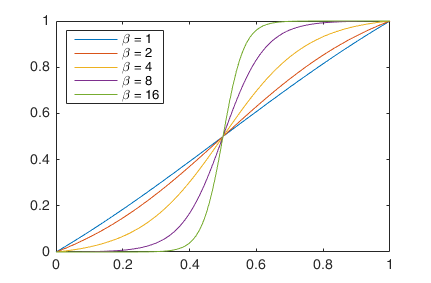}
\caption{The projection function Eq.~\ref{eq:softBlackWhiteFilter} for various $\beta$ values. As $\beta$ increases, the function approaches the discrete function Eq.~\ref{eq:blackWhiteFilter}.}
\label{fig:beta}
\end{figure}

\noindent \textbf{Projection $\tilde{\phi} \to \rho$} \quad 
The purpose of the projection $\tilde{\phi} \to \rho$ is to ensure a 0-1 solution. An intermediate value between $0.0$ and $1.0$ is thresholded at the value of $\frac{1}{2}$ to a discrete 0/1 value by
\begin{equation}
\rho_e (\tilde{\phi}_e) = \begin{cases}
1 & \quad \text{if $\tilde{\phi}_e \ge \frac{1}{2}$,} \\
0 & \quad \text{otherwise}.
 \end{cases}
\label{eq:blackWhiteFilter}
\end{equation}
For numerical optimization, we relax $\rho_e$ to a scalar threshold function, and approximate this non-differential function by
\begin{equation}
\rho_e (\tilde{\phi}_e) = \frac{\tanh(\frac{\beta}{2})+\tanh(\beta(\tilde{\phi}_e-\frac{1}{2}))}{2\,\tanh(\frac{\beta}{2})} .
\label{eq:softBlackWhiteFilter}
\end{equation}
The parameter $\beta$ controls the sharpness of the threshold function, as illustrated in Fig.~\ref{fig:beta}. An infinite $\beta$ leads to a strict binary classification as in Eq.~\ref{eq:blackWhiteFilter}. Instead of directly applying a large $\beta$ value, which results in highly non-linear equations, we start with $\beta=1$ and double its value after a certain number of iterations. This process is known as parameter continuation, which is a common technique for improving convergence behaviour~\cite{Wang10}.

\noindent \textbf{Material Interpolation} \quad With $\rho_e$ being relaxed via Eq.~\ref{eq:softBlackWhiteFilter} to a scalar, the Young's modulus corresponding to a voxel with continuous material distribution $\rho_e$ is interpolated by
\begin{equation}
E_e(\rho_e) = E_{min} + \rho^{\gamma}_e \, (E_0 - E_{min}).
\label{eq:YoungsModulus}
\end{equation}
Here $E_0$ is the stiffness of the solid voxels, $E_{min}$ is a very small stiffness assigned to empty voxels, in order to prevent the global stiffness matrix from becoming singular, and $\gamma$ is a penalization factor (typically $\gamma=3$). Assuming a fixed Poisson's ratio, the stiffness matrix of intermediate voxels then becomes
\begin{equation}
K_e = E_e(\rho_e)k_0,
\label{eq:elementStiffness}
\end{equation}
where $k_0$ is the element stiffness matrix for a voxel with unit Young's modulus. This interpolation scheme is known as the modified Solid Isotropic Material Penalization (SIMP)~\cite{Andreassen10}.

\subsection{Relaxed Formulation}
With the above relaxations, the optimization problem becomes
\begin{align}
\underset{\phi}{\text{min}}
&\quad c = \frac{1}{2} u^T K u, \label{eq:continuousObjective} \\
\text{s.t.}
&\quad Ku = f,  \\
&\quad \phi_e \in [0.0,\;1.0],  \; \forall e, \label{eq:continuousDesign} \\
&\quad g(\phi) = \frac{ \textstyle{ ( \frac{1}{n} \sum_{e} \overline{\rho}^p_e ) ^{\frac{1}{p}} } }{\alpha} - 1.0 \le 0.0. \label{eq:aggregatedConstraint}
\end{align}
Here, the design variable is the continuous variable $\phi$. This continuous optimization problem is solved iteratively by using gradient-based optimization schemes. In each iteration, three major steps are performed sequentially:
\begin{enumerate}
\item solve the state equation $Ku=f$ for the unknown displacement vector $u$,
\item do sensitivity analysis to get the derivatives of the objective and the constraint function with respect to the design variable $\phi$, i.e., $\frac{\partial c}{\partial \phi}$ and $\frac{\partial g}{\partial \phi}$, and 
\item update the design variables by a numerical optimization solver.
\end{enumerate}
These three steps continue until the change of design variables in successive iterations falls below a prescribed threshold $\epsilon$, or the number of iterations exceeds a maximum value $It_{\max}$. For the numerical optimization solver in step 3) we use the method of moving asymptotes (MMA)~\cite{Svanberg87,Aage13}. 

We detail the optimization process in Algorithm~\ref{alg:optimizationRoutine}. The algorithm takes as input the prescribed local volume fraction $\alpha$, and outputs the density field $\rho$ which represents the material distribution. 

\begin{algorithm}
\caption{\label{alg:optimizationRoutine}Infill optimization}
\begin{algorithmic}[1]
 \renewcommand{\algorithmicrequire}{\textbf{Input:}}
 \renewcommand{\algorithmicensure}{\textbf{Output:}}
 \renewcommand\algorithmiccomment[1]{\hfill $\vartriangleright$ \eqparbox{COMMENT}{#1}}
\REQUIRE Local volume fraction $\alpha$
 \ENSURE  Density field $\rho$
 \\ \STATE Design variable $\phi=\alpha$
 \\ \STATE Iteration index $i = 0$
 \\ \STATE Design change $\Delta = 1.0$ 
 \\ \STATE Projection parameter $\beta = 1.0$ 
 \WHILE{$\Delta > \epsilon$ and $i \le It_{\max}$}
	\STATE $i = i+1$
    \STATE $\tilde{\phi} \leftarrow \phi $ via Eq.~\ref{eq:densityFilter} 
    \STATE $\rho \leftarrow \tilde{\phi}$ via Eq.~\ref{eq:softBlackWhiteFilter} 
    \STATE $K \leftarrow \rho$ via Eq.~\ref{eq:YoungsModulus} \& \ref{eq:elementStiffness} 
    \STATE $u$ via solving $Ku=f$
    \STATE $c \leftarrow \left(u, K\right)$ via Eq.~\ref{eq:continuousObjective}
    \STATE $\overline{\rho} \leftarrow \rho$ via \revise{Eq.~\ref{eq:localMeasure}}{Eq.~\ref{eq:localDensity}}
    \STATE $g \leftarrow \left(\overline{\rho}, \alpha \right)$ via Eq.~\ref{eq:aggregatedConstraint}
    \STATE $\frac{\partial c}{\partial \phi}$ \& $\frac{\partial g}{\partial \phi}$ as in Appendix
    \STATE $\phi \leftarrow \left( c, g, \frac{\partial c}{\partial \phi}, \frac{\partial g}{\partial \phi} \right)$ via the MMA solver~\cite{Aage13}
    \STATE $\Delta=\max\limits_{\forall e}(|\phi^i_e-\phi^{i-1}_e|)$
    \STATE \textbf{if} $\operatorname{mod}(i,40)==0$ or $\Delta < \epsilon $ \textbf{then}
    \STATE \quad $\beta = 2\beta$ 
    \STATE \quad $\Delta = 1.0$ 
    \STATE \textbf{end if} 
\ENDWHILE
\STATE Compute $\phi \to \tilde{\phi} \to \rho$ 
\end{algorithmic}
\end{algorithm}

\begin{figure*}[t]
\centering
\includegraphics[width=0.98\linewidth]{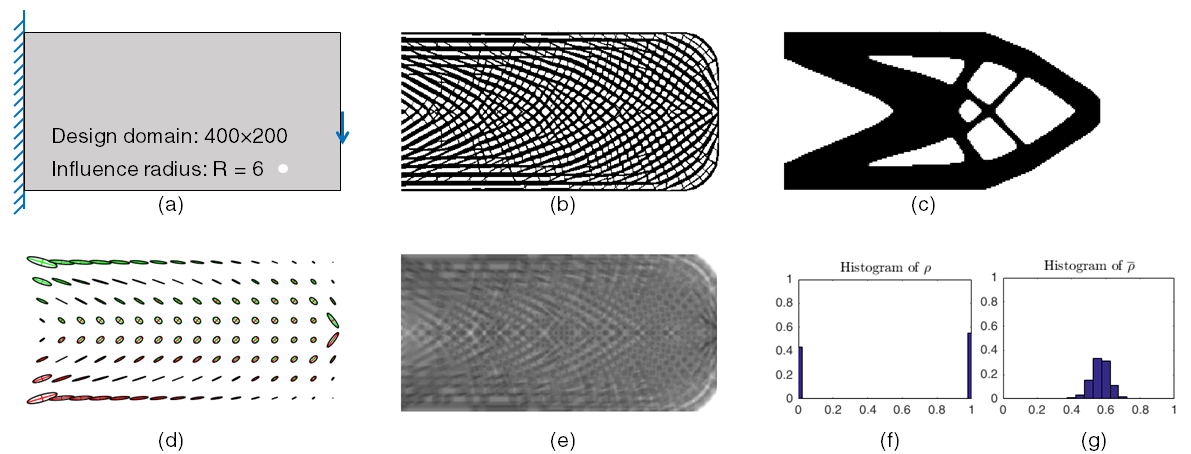}
\caption{(a) Illustration of the design domain, boundary conditions, and the local volume measurement region indicated by the size of the white disk. (b) Structure optimized by the proposed topology optimization with local volume constraints, i.e., the scalar field of $\rho$. (c) Structure optimized by classical topology optimization with a total volume constraint. The compliance of (b) and (c) is $76.86$ and $57.13$, respectively, meaning that the structure considering local volume constraints is somewhat less stiff. (d) Visualization of the stress tensor field in the initial solid design domain. (e) The scalar field of local volume fraction $\overline{\rho}$. (f) The histogram of the obtained material distribution $\rho$. The values converge to a 0-1 solution. (g) The histogram of the local volume fraction $\overline{\rho}$. Most values fall below the prescribed local volume limit $0.6$. 
}\label{fig:cantilever}
\end{figure*}

\subsection{Example}

To demonstrate the effects of the proposed changes to classical topology optimization, a simple 2D example is used in the following. Fig.~\ref{fig:cantilever} (a) shows a rectangular 2D design domain. The left edge of the design domain is fixed, \revise{and o}{meaning that the displacements of the vertices along this edge are constrained to zero. O}n the right edge an external force is applied to the mid point. The design domain is discretized into a $400\times200$ uniform grid. A local volume fraction of $\alpha = 0.6$ and an influence radius of $R=6$ are prescribed. 

Fig.~\ref{fig:cantilever} (b) shows the optimized structure, where black indicates solid elements and white indicates empty elements. This structure has several distinctive features. First, compared to classical topology optimization with a prescribed volume constraint (Fig.~\ref{fig:cantilever} (c)), the material distribution does not evolve towards large solid and empty parts. The reason is that in the proposed formulation at most $60\%$ of all voxels in each local neighbourhood are set to the solid state.

Second, the structure is dominated by crossing elongated sub-structures. These sub-structures largely follow principal stress directions as shown in Fig.~\ref{fig:cantilever} (d) where the stress tensor field in the initially solid design domain is visualized via ellipsoidal glyphs. The axes of the ellipses encode the principal stress directions and magnitudes at every voxel center. In addition, the colour of the axes indicate compression (red) or tension (green).
While crossing sub-structures appear almost everywhere in the domain, single separated elongated structures can be found at the top and bottom of the left boundary. In theses regions the stresses are highly anisotropic (Fig.~\ref{fig:cantilever} (d)), and the material distribution has evolved primarily along the largest principal stress direction.  

Third, the material is distributed across the entire design domain. This results from the objective to minimize compliance. If the constraint on material volume is not enforced, the minimization of compliance leads to a completely filled solid. Since local volume constraints are imposed, the optimizer tends to place material at every location up to the maximum allowed volume ($60\%$ in this case). The local volume values $\overline{\rho}$ is shown in Fig.~\ref{fig:cantilever} (e), and the histogram showing the frequency of occurrence of values is given in Fig.~\ref{fig:cantilever} (g). It can be seen that the majority of local volume values is below the prescribed limit $0.6$. The $p$-norm approximation does not represent the $\max$ function accurately. At a few places the prescribed local volume limit is exceeded. These values are mostly located in regions where the stress is very large.

\begin{figure*}[t]
\centering
\includegraphics[width=0.98\linewidth]{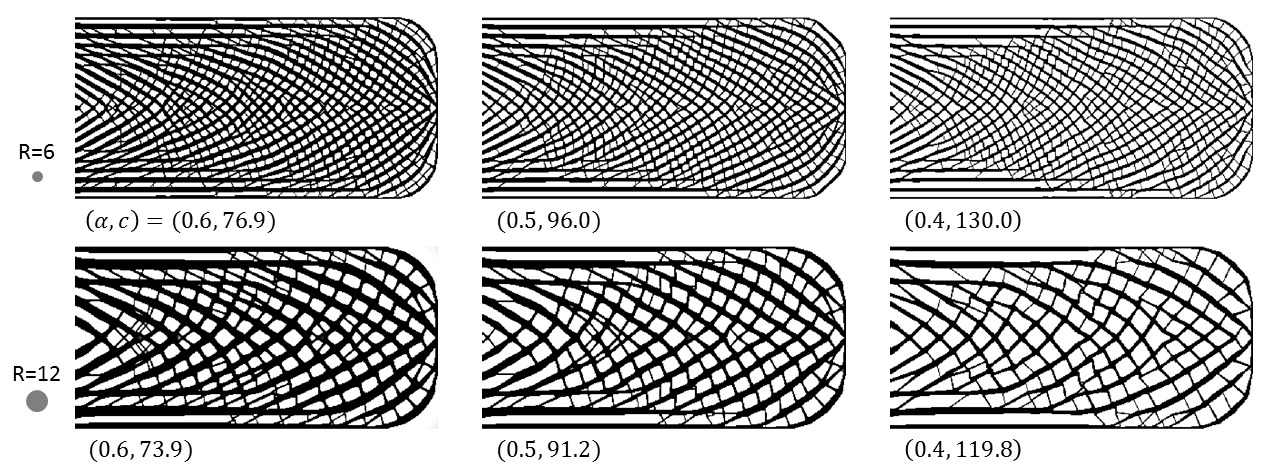}
\caption{\revise{}{From left to right, the local volume limit $\alpha$ deceases from $0.6$ to $0.4$. This leads to increasing porosity and a natural increase in the compliance value thereof. Structures in the bottom row were generated using a larger influence radius $R$, resulting in a less strict locality constraint and increasing stiffness of the structures thereof. In all examples, the same boundary conditions as illustrated in Fig.~\ref{fig:cantilever} (a) were applied.}
}\label{fig:localVolume}
\end{figure*}

\revise{}{The local volume constraint is parametrized by two values, the local volume limit $\alpha$, and the influence radius $R$. The effect of both parameters on the optimized structure is examined in Fig.~\ref{fig:localVolume}. It can be seen that the local volume limit controls the local porosity, while the influence radius controls the empty space between substructures. As the influence radius increases, the locality constraint becomes less strict, leading to stiffer structures. If the influence radius becomes larger than the size of the design domain, the local volume constraints become equivalent to a total volume constraint, resulting in the stiffest structure as in classical topology optimization.}
\section{Extensions}
\label{sec:extension}
To provide further control over the optimized structures, we present several extensions to the infill optimization formulated by Eq.~\ref{eq:continuousObjective}-\ref{eq:aggregatedConstraint} (referred to as the basic formulation in the following), and analyze the resulting infills in comparison to those resulting from the basic formulation.

\begin{figure*}[t]
\centering
\includegraphics[width=0.98\linewidth]{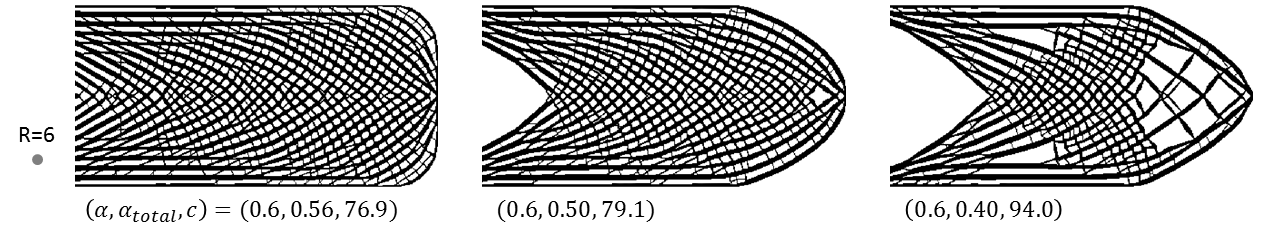}
\caption{\revise{}{In addition to a local volume limit of $\alpha=0.6$, limits of $\alpha_{total} = 0.5$ (middle) and $\alpha_{total} = 0.4$ (right) were imposed on the global material distribution. The compliance increases due to the use of less material. The applied boundary conditions are shown in Fig.~\ref{fig:cantilever} (a).}
}\label{fig:totalVolume}
\end{figure*}

\subsection{Total Volume Control}
By prescribing a maximum value for the total material volume, the user can control the expected cost of a print. Representing the voxel volume by $v_e$, which is constant in the regular discretization, the solid volume normalized by the volume of the design domain is 
\begin{equation}
\rho_{avg} = \frac{\sum\limits_e\rho_e v_e}{\sum\limits_e v_e}.
\end{equation}

The local volume constraint $g$ (Eq.~\ref{eq:aggregatedConstraint}) implicitly imposes an upper bound on the total volume. In fact, the local upper bound $\alpha$ is a good indicator for the ratio of total volume $\rho_{avg}$, i.e., if $\rho_e = \alpha, \forall e$, we get $\rho_{avg} = \alpha$. However, since the density values $\rho_e$ are set to either 0 or 1, and due to the domain boundaries, the resultant $\rho_{avg}$ is smaller than $\alpha$. For instance, in the 2D test example, $\alpha = 0.6$ leads to $\rho_{avg} = 0.56$. 

To support direct control over the total volume, we integrate the following total volume constraint into the optimization problem:
\begin{equation}
g_1 = \rho_{avg} - \alpha_{total} \le 0.0,
\end{equation}
where $\alpha_{total}$ is a user-selected limit on the total volume ratio. The integration of this constraint into Algorithm~\ref{alg:optimizationRoutine} is straightforward. The constraint value $g_1$ and the value of its derivative $\frac{\partial g_1}{\partial \phi}$ are calculated and fed into the optimizer, together with their counterparts controlling the local volume, i.e., $g$ and $\frac{\partial g}{\partial \phi}$.

Fig.~\ref{fig:totalVolume} shows the result when different limits on the total material volume are used for optimizing the infill of the test shape in Fig.~\ref{fig:cantilever} (a). In all cases, a local volume limit $\alpha = 0.6$ is imposed. In Fig.~\ref{fig:totalVolume}  (middle), as the total volume is controlled, structures disappear in regions of lower stresses (cf. the stress visualization in Fig.~\ref{fig:cantilever} (d)). In Fig.~\ref{fig:totalVolume}  (right), as the total volume is further reduced, the material distribution shrinks from low stress regions and evolves towards the structures that are generated by classical topology optimization (cf. Fig.~\ref{fig:cantilever} (c)).

\begin{figure*}[t]
\centering
\includegraphics[width=0.98\linewidth]{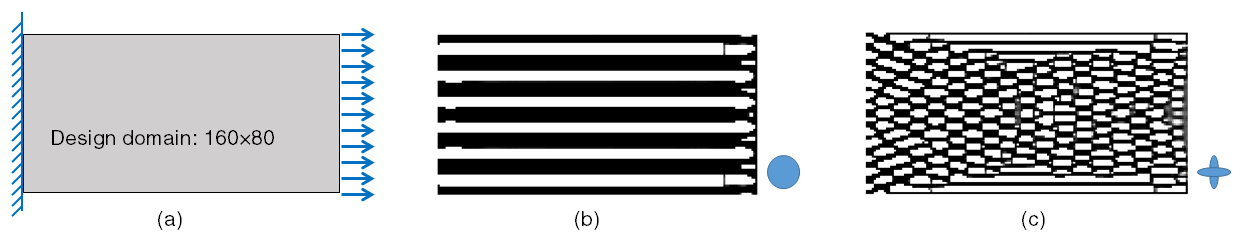}
\caption{Comparison of 2D structures optimized with isotropic and anisotropic filters. (a) Illustration of the design domain and boundary conditions. (b) The structure optimized with an isotropic filter, the size of which is indicated by the blue disk. (c) The structure optimized with anisotropic filters. The compliance and the total volume of (b) and (c) are $22.6$ with $59.8\%$ volume, and $34.6$ with $51.7\%$ volume, respectively.
}
\label{fig:2DIso-aniso}
\end{figure*}

\begin{figure}[t]
\centering
\includegraphics[width=0.6\linewidth]{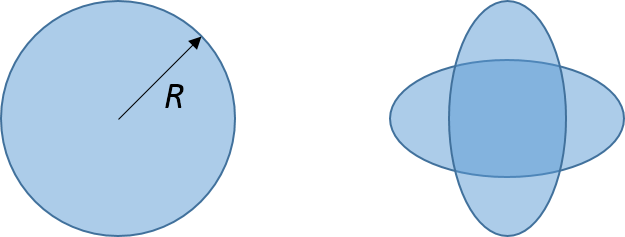}
\caption{Illustration of the influence region of a 2D isotropic filter (left) and two orthogonal anisotropic filters (right).}
\label{fig:anisotropic}
\end{figure}

\begin{figure*}[t]
\centering
\includegraphics[width=0.88\linewidth]{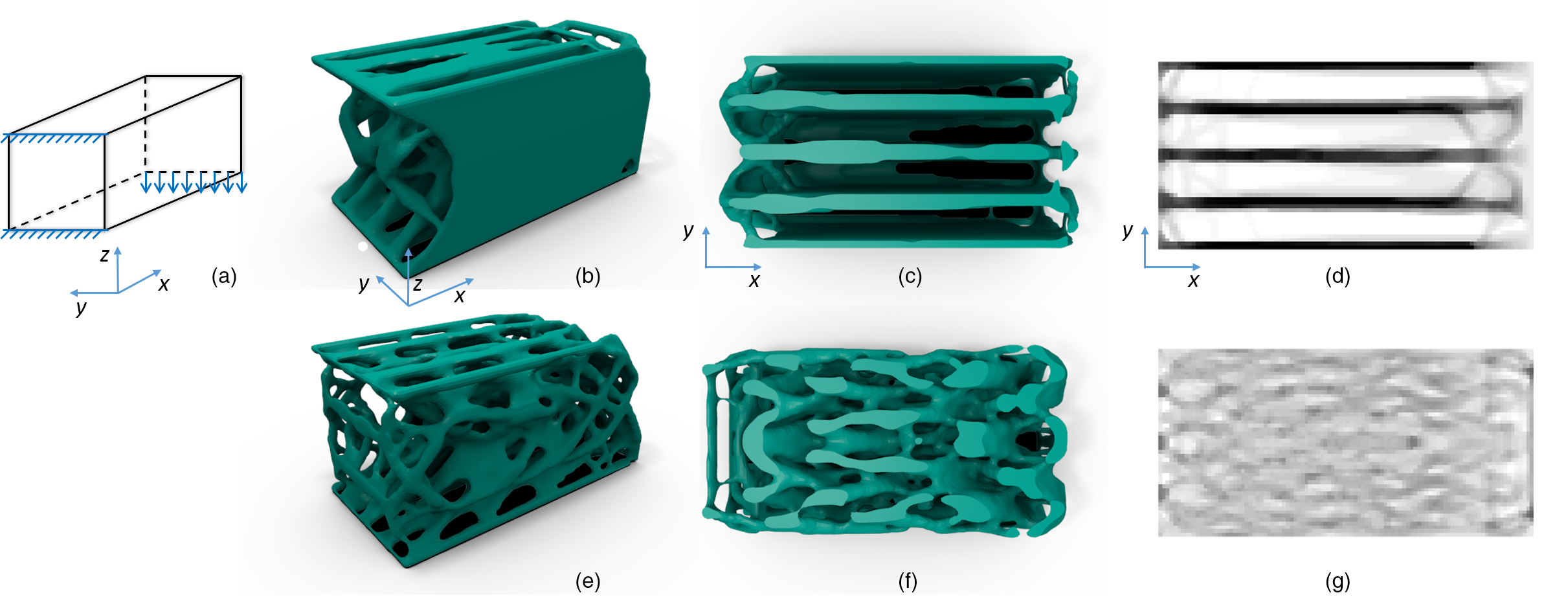}
\caption{Comparison of 3D structures optimized with isotropic and anisotropic filters. (a) Illustration of the design domain and boundary conditions. (b) The  structure optimized with an isotropic filter. (c) The structure in (\revise{a}{b}) viewed from top, with a cut-plane parallel to the $xy$-plane. (d) The 3D density field is projected \revise{i}{o}nto a 2D $xy$-plane\revise{}{, visualizing the distribution of averaged densities along rays parallel to the $z$-axis}. (e,f,g) The structure optimized with anisotropic filters. The compliance and the total volume resulting from isotropic and anisotropic filters are $79.4$ with $27.9\%$ volume, and $125.6$ with $23.8\%$ volume, respectively.
}
\label{fig:3DIso-aniso}
\end{figure*}

\subsection{Anisotropic Filter}
In the basic formulation, we define the local volume fraction in a circular neighbourhood, and treat all elements in this neighbourhood the same.
However, since the stress distribution at some locations can be highly anisotropic, in such regions the material will accumulate along the major principal direction, while leaving the other direction weakly or barely connected. This can be seen in the top and bottom left parts of the 2D object shown in Fig.~\ref{fig:cantilever}-\ref{fig:totalVolume}.
To further demonstrate the effect of anisotropy in the stress distribution, we show in Fig.~\ref{fig:2DIso-aniso} (a) a situation where the left edge of the cantilever is fixed, while uniformly distributed horizontal forces are applied to the right edge. Due to high uni-axial tension along the horizontal direction, the optimized structure is almost solely composed of horizontal bars (see Fig.~\ref{fig:2DIso-aniso} (b)).  

To distribute the material along all directions, and thus to simulate natural bone remodelling, we suppress unidirectional growth by using anisotropic filters for defining the local volume fractions. \revise{As we don't know the principal directions \textit{a priori}, w}{} Two and three such filters are used in 2D and 3D, respectively, with a 2D example shown in Fig.~\ref{fig:anisotropic}. \revise{}{Here, $90^{\circ}$ degree orientations were used, yet other configurations can be used as well. For instance, one could use $60^{\circ}$ oriented filters to obtain a higher degree of isotropy, or the filter axes could be oriented automatically along the principle stress direction determined by the finite element analysis of the initial shape.} In 3D, the isotropic local volume measure $\overline{\rho}_e$ is substituted by three local volume measures corresponding to three different filter orientations,
\begin{equation}
\overline{\rho}_{e,s} = \frac{\sum_{i\in \mathbb{N}_{e,s}} \rho_i}{\sum_{i\in \mathbb{N}_{e,s}} 1}, \quad s\in\{x,y,z\},
\end{equation}
where $\mathbb{N}_{e,s}$ is the set of elements in the anisotropic influence region. Consequently, the constraint $\overline{\rho}_e \le \alpha, \forall e$ is replaced by
\begin{equation}
\overline{\rho}_{e,s} \le \alpha, \forall e, s\in\{x,y,z\}.
\end{equation}

Fig.~\ref{fig:2DIso-aniso} (c) shows the resulting structures when optimizing the structure in (a) using two anisotropic filters. It can be seen clearly that the horizontal bar-like structures are broken up, and instead the optimization process tries to connect short horizontal and vertical sub-structures.

In Fig.~\ref{fig:3DIso-aniso}, the effects that can be achieved by using anisotropic filters in 3D are demonstrated. When the isotropic filter is used (top row), there are only a few connections along the $y$-axis, since the stresses in the $xz$-plane are larger than those along the $y$-axis. When anisotropic filters are used (bottom row), more bridge-like connections between the planes parallel to the $xz$-plane are generated.

\subsection{Truss- vs. Wall-like Structures}
\label{subsec:truss-vs-wall}
The local volume constraint suppresses the emergence of large solid domains. Two types of sub-structures can emerge primarily from such a constraint, thin walls and trusses. A mix of both types is visible in 3D examples (see Fig.~\ref{fig:femur}). In the following, we discuss the parameters used to control the type of sub-structures, and propose reformulations to prioritize or suppress certain types. 

It has been shown recently~\cite{Sigmund16}, that thin-walls are the most effective 3D structures for stiffness optimization, as opposed to truss-like structures. Nevertheless, truss-like structures do appear very often in structural optimization, which is primarily due to an insufficient resolution of the underlying simulation grids. The spatial discretization implicitly requires the walls, if they exist, not to be thinner than one simulation element (and more if filtered). If the local material allowance is not sufficient for creating wall-like structures, holes are formed in the (\revise{yet not}{not yet} developed) walls. Given a higher resolution discretization, wall-like structures connecting the truss-like structures from a low spatial discretization emerge.

While from the perspective of solid mechanics closed-walled structures are more optimal, truss-like structures are found to dominate in trabecular bone. This can be attributed to the involved biofluid mechanics~\cite{Birmingham13}, i.e., truss-like structures allow unblocked interaction between the solid structures and the surrounding fluid environment. Truss-like structures are also preferable for their superior manufacturability, i.e., reducing the possibility of trapped powders in the post-processing of printed models. Since wall-like structures below the minimum feature size will fall apart into truss-like structures, the idea is to prescribe a larger minimum feature size (or alternatively a smaller maximum local volume) and, thus, to explicitly enforce the breakdown of closed-walled structures. The minimal feature size can be controlled by the filter radius $r$ in the projection $\phi \to \tilde{\phi}$, as thoroughly studied in~\cite{Guest04,Wang10}.

\begin{figure}[t]
\centering
\includegraphics[trim = 0mm 25mm 0mm 15mm, clip, width=0.98\linewidth]{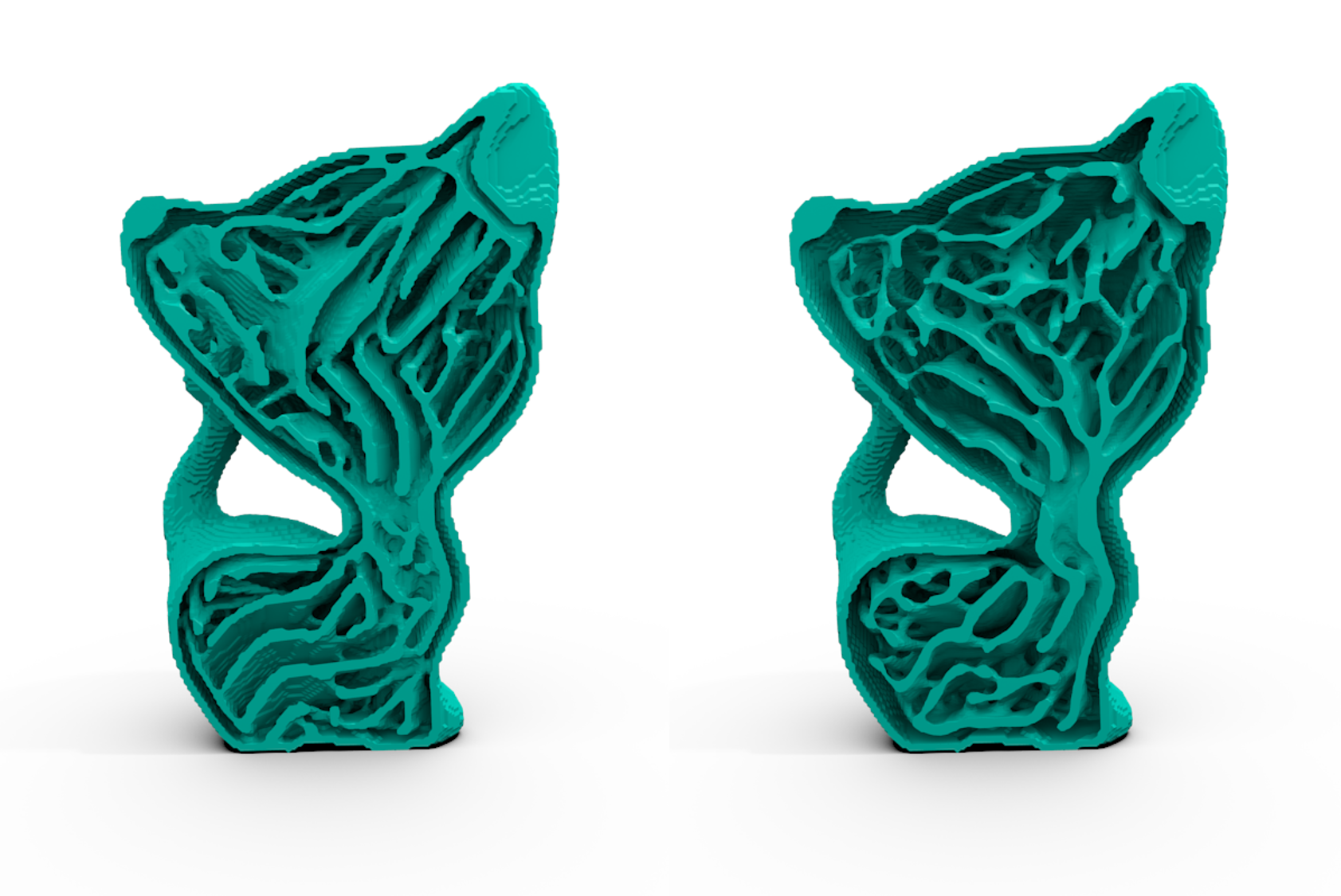}
\caption{
As the minimum thickness increases from left to right, more truss-like structures appear, replacing wall-like structures.
}\label{fig:truss-vs-wall}
\end{figure}

Fig.~\ref{fig:truss-vs-wall} compares the optimized structures with different minimal feature sizes: $r = 2$ (left), and $r=3$ (right). It can be seen that as the feature size increases, some walls are substituted by a sparse set of trusses. Along with the breakdown of closed-walled structures, we observe a decrease of stiffness by a factor of up to $20\%$.

The relation between the minimal feature size and the structural types can be derived analytically. Consider a location $x_e$ in the design domain. The influence region of an isotropic filter with radius $R$ has a volume of $V_{sphere} = \frac{4}{3}\pi R^3$. A wall with a thickness of $2r$ takes a volume of $V_{wall} = 2 \pi rR^2 - \frac{2}{3}\pi r^3$. To suppress the emergence of the wall, the allowed volume ratio $\alpha$ should be smaller than the required volume fraction, i.e., 
\begin{equation}
\alpha < \frac{V_{wall}}{V_{sphere}}.
\end{equation}
This equation leads to the lower bound of the radius $r$ under a prescribed volume allowance of $\alpha$, and interchangeably, the upper bound of volume ratio $\alpha$ under a prescribed feature size $r$.

The above analysis is based on the assumption of strictly enforced local volume constraints, which is computationally prohibitive to realise. In our implementation, since we approximate the local volume constraints by a global $p$-norm, the actual volume fraction at locations where the stress is extremely high can be larger than the prescribed limit. Consequently, at such locations wall-like structures may still emerge.

\subsection{Passive Elements}
To fix a thin shell below the surface of an input 3D model, we prescribe elements close to the surface mesh as \textit{passive}, and denote the remaining elements as \textit{active}. To this end, we compute a distance field in the design domain $\Omega$, representing the shortest distance from the centroid of each element to the surface mesh. Elements with distances that fall below a prescribed layer thickness $t$ are identified as passive. The thickness value can be adjusted by the user. In our examples, we typically prescribe a thickness of 2 in terms of voxels.

The passive elements are excluded from the design update step -- The passive elements maintain $\rho_e = 1.0$. These solid elements are considered in the finite element analysis, since they can sustain forces as well. Passive elements are also excluded from the calculation of local volume fraction for active elements. This is realised by augmenting the set of neighbouring elements, $\mathbb{N}_e$ in Eq.~\ref{eq:localMeasure}, by
\begin{equation}
\mathbb{N}_e=\{i| \; ||x_i-x_e||_2 \le R_e, i \not\in \Omega_s \},
\end{equation}
where $e$ refers to an active element, and $\Omega_s$ is the set of passive elements.

\section{Results and Analysis}
\label{sec:results}

We have implemented the proposed infill optimization method in 2D based on the Matlab code provided in~\cite{Andreassen10}, and in 3D based on the high-performance multigrid solver for topology optimization detailed in~\cite{Wu16}. \revise{}{In 3d, a surface mesh is constructed from the optimized scalar density field via the Marching Cubes algorithm~\cite{Lorensen87} in a post-process, and this mesh is smoothed via Taubin smoothing~\cite{Taubin95} to eliminate staircase artefacts.} In the following, we present and discuss a number of additional infills that have been generated by our method.

\begin{figure}[t]
\centering
\includegraphics[width=0.98\linewidth]{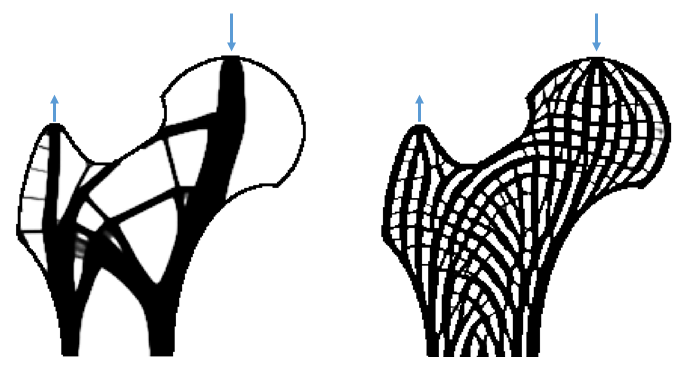}
\caption{Topology optimization in a 2D femur-shaped design domain. Left: Classical topology optimization with a total volume constraint. Right: Proposed topology optimization with local volume constraints.
}\label{fig:2DBone}
\end{figure}

\begin{figure}[t]
\centering
\includegraphics[width=0.8\linewidth]{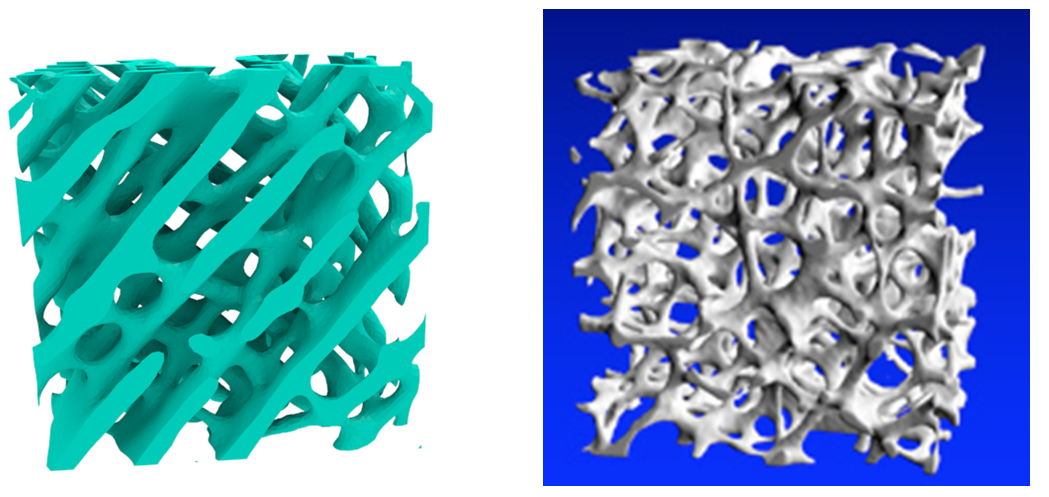}
\caption{Visual comparison of a cubic sample taken from the optimized infill in Fig.~\ref{fig:femur} and a real bone sample from CT scans (Image courtesy of R. M\"uller ~\cite{Muller97}).
}\label{fig:boneSample}
\end{figure}

\subsection{Bone Models}
To verify that the proposed formulation leads to infills with a similar structure than trabecular bone, we prescribe a 2D femur-shaped design domain as shown in Fig.~\ref{fig:2DBone}, and run the optimization with a total volume constraint (left), and local volume constraints (right). The porous infill on the right clearly follows principal stress directions as depicted in Fig.~\ref{fig:femur}.

As already indicated by Fig.~\ref{fig:femur}, the 3D results are also very promising. The femur model is simulated with a resolution of $280 \times 185 \times 364$, leading to a total of $5.56$ million finite elements. \revise{The optimization process takes about 6 hours.}{} The femur model with the optimized infill was fabricated by using selective laser sintering. The selected material is a strong flexible plastic. The physical replica has a dimension of $12.32\,cm \times 5.85\,cm \times 16.00\,cm$.

Fig.~\ref{fig:boneSample} compares a cubic sample taken from the optimized infill (left) and a sample from a human femur CT scans (right). It can be seen that both samples are composed of sparse trusses and a few walls. 

\subsection{Robustness}

\begin{figure*}[t]
\centering
\includegraphics[width=0.98\linewidth]{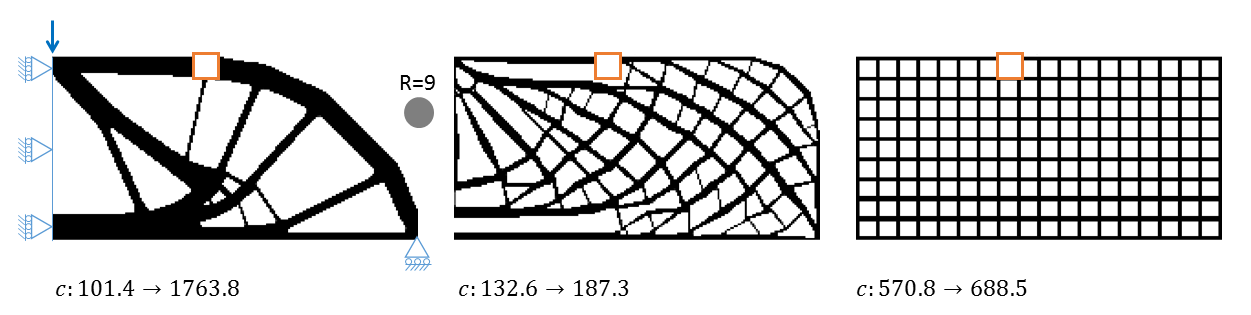}
\caption{
{\revise{}{Illustration of local material damage. Without damage, the structure constrained to a total volume (left) has the smallest compliance $c$ of the three structures, but when damaged as indicated by the orange square its compliance becomes worst. The middle structure was generated by constraining the local volume limit. A regular grid structure (right) serves as reference.}}
}\label{fig:damage}
\end{figure*}

\begin{figure}[t]
\centering
\includegraphics[width=0.9\linewidth]{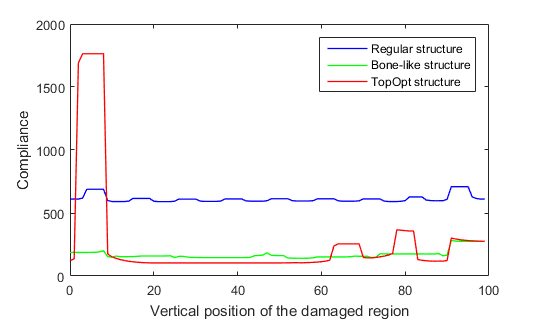}
\caption{
{\revise{}{Change in compliance with respect to a constantly relocated damage region for the three designs in Fig.~\ref{fig:damage}. The bone-like porous infill (green) has a smaller variation, and a small worst-case compliance.}}
}\label{fig:damageCurve}
\end{figure}

\noindent \textbf{With respect to material deficiency} \quad 
An advantage of distributed porous structures is their damage tolerance, i.e., the infill is still stiff even if parts are broken. In practice, the structures are subject to local damage, for instance, due to accidental collisions or manufacturing errors. \revise{}{To simulate the effect of damage on a structure's compliance, we employ a simplified local damage model~\cite{Jansen14}. Here it is assumed that a quadrilateral region of a fixed size is damaged, and that this region can be placed everywhere in the design domain. It is desired that in case of damage a high stiffness can be maintained.} 

\revise{}{We test the damage tolerance of different structures generated for the half MBB-beam at a grid resolution of $200\times100$ (see Fig.~\ref{fig:damage}). We first optimize with respect to local volume constraints using a local volume limit of $\alpha=0.4$ (middle). The influence radius is selected based on the assumed damage size, effectively controlling the amount of empty space between porous structures. The resulting volume ($\alpha_{total} = 0.368$) is then considered in a second optimization with respect to the total volume constraint (left). A regular grid (right) with the same volume serves as a reference. To exactly match the prescribed volume, the thickness of the horizontal bars is slightly enhanced at the bottom.}

\revise{}{To simulate damage, we remove the material in a certain region---indicated by the orange squares in Fig.~\ref{fig:damage}---from the optimized structures. The damaged shapes are then compared with respect to their compliance. While the total volume constrained infill (left) has a compliance of $101.4$ before and $1763.8$ after damage, the local volume constrained infill (middle) has a compliance of $132.6$ before and $187.3$ after damage. This suggests that the infill that was optimized with respect to the total volume is very sensitive to material damages -- The compliance changes by a factor of $17.4$, while it changes only by a factor of $1.4$ for the porous infill. }

\revise{}{To consider the sensitivity of the structural compliance to the applied damage, we vertically move the damage region downwards and consecutively evaluate the compliance. The curves in Fig.~\ref{fig:damageCurve} show the resulting changes for all three test structures. It can be seen that the bone-like infill (green curve) exhibits only small variations in the compliance values. The total volume constrained structure undergoes large changes under some damage conditions. The regular infill, even though it also shows only minor variations, the compliance values are about $4$ times larger than those of the bone-like infill.}

\begin{figure*}[t]
\centering
\includegraphics[width=0.88\linewidth]{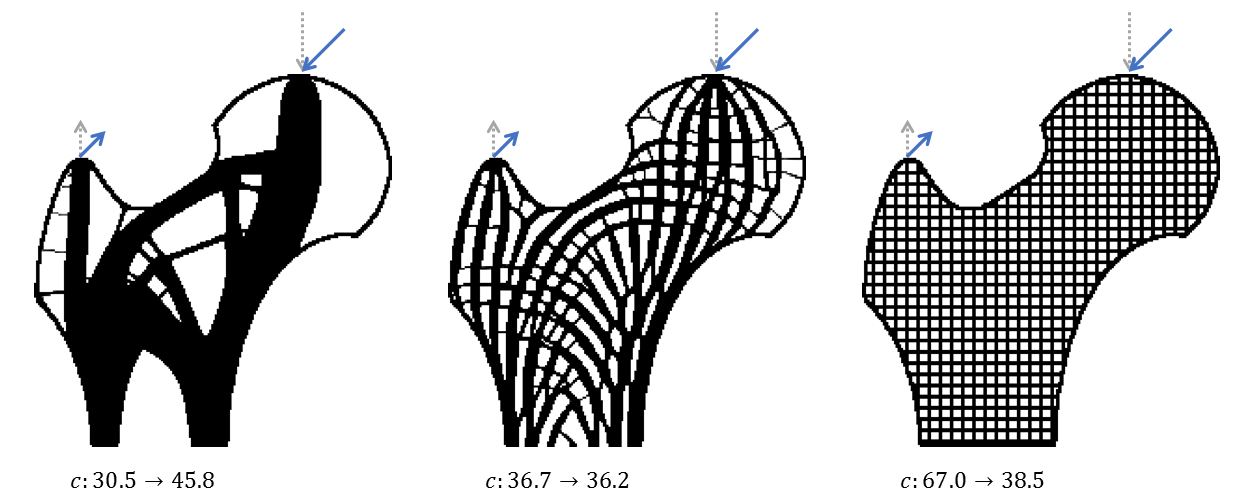}
\caption{
{\revise{}{The structures are optimized with respect to the force condition as indicated by the dashed grey arrows. Under alternative forces (solid blue arrows), the local volume constrained structure (middle) is $1.3$ times stiffer than the total volume constrained structure (left). The regular grid with the same amount of volume on the right serves as a reference.}}
}\label{fig:robustness}
\end{figure*}

\begin{figure}[t]
\centering
\includegraphics[width=0.9\linewidth]{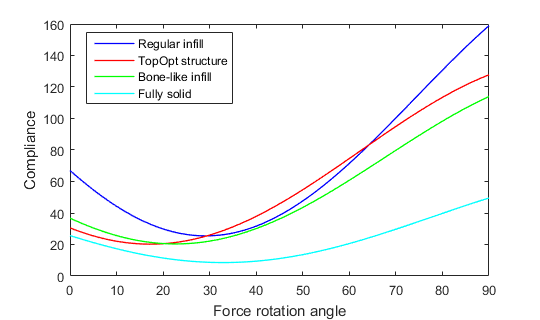}
\caption{
{\revise{}{The compliance with respect to changed force directions for the 2D femur model (Fig.~\ref{fig:robustness}). The bone-like porous infill (green) is less sensitive to large changes in the force direction, and has a smaller worst-case compliance compared to the total volume constrained structure (red).}}
}\label{fig:forceVariationCurve}
\end{figure}

\noindent \textbf{With respect to force variations} \quad 
The second benefit of porous structures is their robustness with respect to force variations. In practical use cases of consumer products, the external forces are often not constant in direction and/or magnitude and are subject to changes. 

We test the structure performance under such uncertain force conditions for the 2D bone example, as illustrated in Fig.~\ref{fig:robustness}. The structures are optimized with respect to the forces indicated by the dashed grey arrows, under a total volume constraint (left) and the local volume constraints (middle). We then rotate the forces by $\frac{\pi}{4}$ as represented by the solid blue arrows, and re-calculate the compliance of the structures under the new force condition. The total volume constrained infill is very sensitive to this change of direction -- The compliance changes from $30.54$ to $45.83$. In contrast, the local volume constrained version only changes from $36.72$ to $36.23$. 

\revise{}{To examine the sensitivity of compliance  to varying force directions, we consecutively rotate the force directions out of an initial start configuration. Fig.~\ref{fig:forceVariationCurve} shows the compliance values depending on the rotation of force direction. For all four infill configurations, the compliance values change, and each forms a valley-shaped curve. The regular infill (blue, top) and the fully solid infill (cyan, bottom) serve as references. At the rotation angle of zero, for which the structures are optimized, both the total volume (red) and local volume (green) constrained structures have a compliance close to that of the solid structure, with the total volume constrained structure performing even better. At rotation angles up to about $20^{\circ}$, the red curve remains below the green one, indicating a somewhat higher stiffness of the total volume constrained structure. Beyond $20^{\circ}$, the red curve exceeds the green one, indicating the superiority of the local volume constrained structure as well as the sensitivity of total volume constrained structure to large directional changes. At the rotation angle of $90^{\circ}$, both structures show their worst-case compliance.}

This test verifies the robustness of the local volume \revise{constraints}{constrained structure}. This effect agrees with the formation of porous structures in bone. In fact, the mechanical load applied to the bone is not static, but varies during our daily lives. The bone undergoes dynamical natural optimization, and produces porous structures accounting for all different load conditions.

\begin{figure*}[t]
\centering
\includegraphics[width=0.9\linewidth]{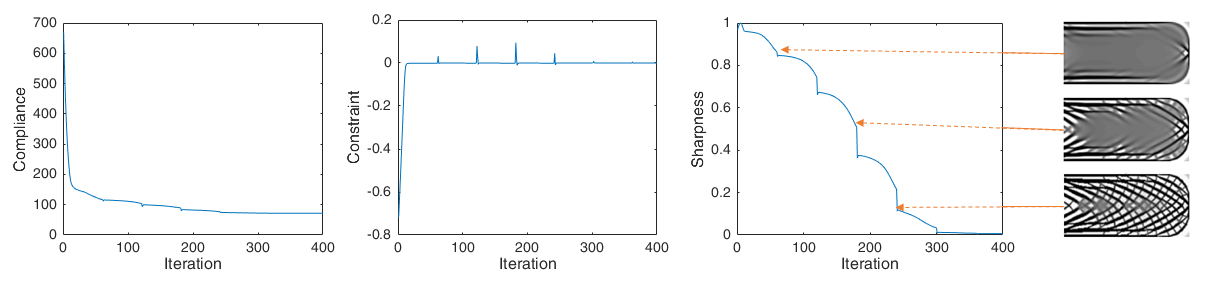}
\caption{Convergence plots of the compliance (left), the constraint (middle), and the sharpness of the density field (right) over the iterative optimization process. The density field, shown at three different stages on the right, gradually converges to a 0-1 solution by using $\beta$-continuation in Eq.~\ref{eq:softBlackWhiteFilter}, which explains the discontinuity in the plots at every $40$ iterations.
}
\label{fig:convergence}
\end{figure*}

\subsection{Convergence \& Performance}

\noindent
\revise{}{\textbf{Convergence} \quad}
We analyze the convergence of the numerical optimization scheme on the 2D cantilever model in Fig.~\ref{fig:convergence}. The horizontal axis is the number of iterations, while the vertical axis (left) represents the compliance value, i.e., the objective in Eq.~\ref{eq:continuousObjective}, and (middle) represents the constraint in Eq.~\ref{eq:aggregatedConstraint}. It can be observed that the compliance value gradually decreases during the optimization process, meaning that the structure becomes stiffer. The constraint value is maintained below $0.0$, while a few jumps happen when the $\beta$ value is doubled every $40$ iterations. This $\beta$-continuation is introduced to convert the intermediate values into a strict 0-1 solution cf. earlier discretization. We define \textit {sharpness} to measure how close the continuous density field is to a binary field~\cite{Sigmund07},
\begin{equation}
s=\frac{4}{n}\sum_e(\rho_e (1-\rho_e)),
\end{equation}
where $n$ is the number of elements. When the density values are converged to a strict 0-1 solution, the sharpness factor becomes $0.0$, while if all elements take a value $0.5$, the sharpness value is $1.0$. The third plot in Fig.~\ref{fig:convergence} shows the evolution of sharpness during the optimization process. On its right, three example structures are displayed,  which correspond to the density field at iterations $79$, $159$, and $279$, respectively. As the optimization progresses, the structure is becoming more discrete. \revise{}{Even though the optimization can be stopped before converging to a discrete design---a common practice when using classical topology optimization in industry---this bears the risk of computing a misinterpreted topology.}

\noindent
\revise{}{\textbf{Performance} \quad
Table~\ref{tab:timings} reports the complexity of the used 3D simulation models as well as timing statistics for different parts of the optimization process considering the local volume constraints. All experiments were run on a standard desktop PC equipped with an Intel Xeon E5-1650 v3 processor (12 cores) running at 3.50\,GHz, 32\,GB of RAM, and an NVIDIA GTX1080 graphics card with 8\,GB memory. We break down the computing time into three parts: FEM, sensitivity analysis and data preparation for MMA, and MMA. The computations involved in all stages consist predominantly of matrix and vector operations, and thus are highly parallelizable. The FEM analysis, which is the performance bottleneck, is accelerated via a highly efficient geometric multigrid solver and GPU parallelization~\cite{Wu16}. The multigrid solver is terminated at a residual reduction of $10^{-4}$. For the sensitivity analysis, we execute the convolution operator with a larger $R$ on the GPU, and the other matrix operators with OpenMP. The MMA solver is parallelized with OpenMP as well, following~\cite{Aage13}. The design optimization is an iterative process using a fixed number of iterations. }

\begin{table}[t]
\footnotesize
\setlength{\tabcolsep}{3pt}
\centering
\begin{tabular}{ llllllll }
	\multirow{2}{*}{Model} & \multirow{2}{*}{Resolution} & \multirow{2}{*}{\# Ele.} & \multicolumn{3}{c}  {Per iteration [s]} & \multirow{2}{*}{\# Iter.} & Total  \\
    \cline{4-6}
	 &  &  & FEM & Sens. & MMA &  & [min] \\ \hline
	Femur & 280$\times$185$\times$364 & 5.6e6 & 6.85  & 5.44  & 2.52  & 500  & 121.2  \\
	Kitten & 218$\times$198$\times$334 & 4.6e6  & 5.45  & 4.54  & 1.91  & 120  & 23.8  \\
	Cantilever & 200$\times$100$\times$100  & 2.0e6  & 1.98 & 2.11 & 0.95 & 120  & 10.1 \\
\end{tabular}
\caption{\revise{}{Performance statistics for different models.}}
\label{tab:timings}
\end{table}

\revise{}{\subsection{Comparison}}
\begin{figure}[t]
\centering
\includegraphics[width=0.98\linewidth]{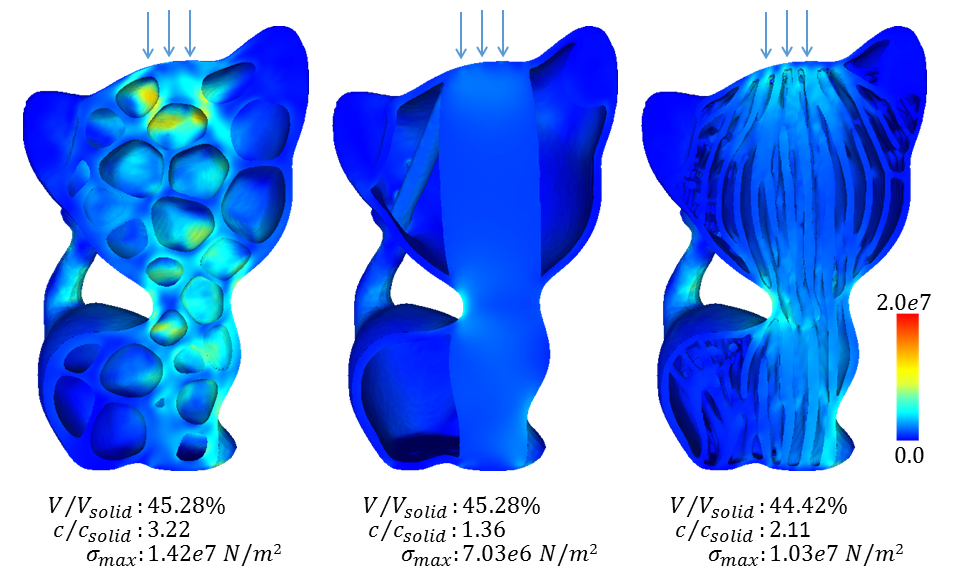}
\caption{
{\revise{}{Comparison between the honeycomb structure~\cite{Lu14} (left, model courtesy of Lu et al.) and structures generated by topology optimization with a total volume constraint~\cite{Wu16} (middle) and with local volume constraints (right).}}
}\label{fig:compareVoronoi}
\end{figure}

\begin{figure}[t]
\centering
\includegraphics[width=0.98\linewidth]{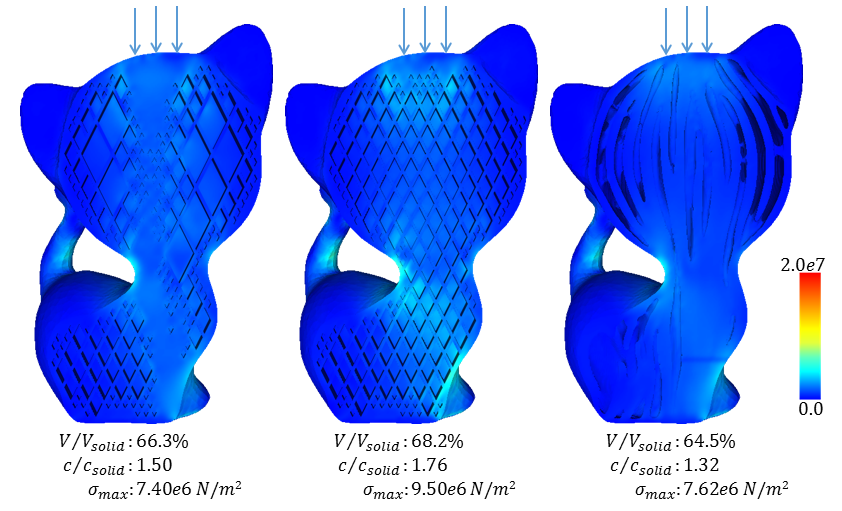}
\caption{
{\revise{}{Comparison between self-supporting rhombic infill structures~\cite{Wu16CAD} that are refined adaptively (left) and uniformly (middle), and the structure generated by local volume constrained topology optimization (right).}}
}\label{fig:compareRhombic}
\end{figure}

\noindent
\revise{}{\textbf{Comparison to honeycomb structures} \quad Fig.~\ref{fig:compareVoronoi} compares the compliance-minimized structures subject to local and global volume constraints to the volume-minimized honeycomb structure subject to a critical stress~\cite{Lu14}. The three models are optimized with the same material properties and boundary conditions. From the stress distribution (left), it can be observed that the forces applied at the top of the kitten model are transmitted mostly through the neck and the tail to the bottom, which is fixed. This results in a straight structure connecting the top and the bottom when using classical topology optimization (middle). Since in the new formulation the local volume is controlled, the vertical structure splits into multiple curved structures (right). The compliance values are normalized against the compliance of a fully solid shape. The values indicate that the porous infill is $1.5$ times stiffer than the honeycomb structure, and has a maximum von Mises stress that is $72.5\%$ of the maximum stress in the honeycomb structure.} 

\noindent
\revise{}{\textbf{Comparison to rhombic structures} \quad In Fig.~\ref{fig:compareRhombic}, we show a comparison to the self-supporting rhombic infill structures~\cite{Wu16CAD}. This example employs the same boundary conditions as in Fig.~\ref{fig:compareVoronoi}, but uses a larger volume percentage in order to allow the rhombic wall to be thicker than three voxels in a stable finite element analysis. The optimized rhombic infill (left) supports the forces by an adaptive refinement in the vertical region. The uniformly refined infill (middle) serves as a reference. The bone-like porous infill (right) exhibits a similar trend as in Fig.~\ref{fig:compareVoronoi} (right). The values show that the porous infill is $1.14$ times stiffer than the optimized rhombic structure, with comparable maximum stresses. Both optimized versions perform better than the uniform grid.}

\revise{}{\subsection{Discussion}}

\noindent
\revise{}{\textbf{Robustness} \quad We have tested bone-like infills under different robustness criteria. Natural materials seem to suggest that structural robustness comes with organized complexity in shape and topology~\cite{Wegst15}. The local volume constraint serves this purpose by encouraging a structural organization of micro-structures to support the prescribed external forces. The very typical approach to ensure robustness to uncertain loads in topology optimization is to optimize with respect to multiple or worst case loading scenarios. This is a wide spread concept (c.f.~\cite{Bendsoe94}, or some quite recent examples~\cite{Zhou13,Allaire09,Schevenels11,Cai15}). However, such approaches require some \textit{a priori} knowledge of the positions of these uncertain loads; anti-optimization problems may have to be solved to identify worst cases; and many load cases (c.f. expensive finite element analyses) must be performed to ensure a reasonable coverage of the uncertainties. To achieve low sensitivity to loading positions and detailed structures as shown in our work, dozens (if not hundreds) of loads with varying location and direction would be necessary. In contrast, our formulation involves only one (or possibly a few) finite element analysis in each iteration. This makes the new formulation more practical for processing the massive models as we see in 3D printing.}

\noindent
\revise{}{\textbf{Manufacturability} \quad While additive manufacturing enables the fabrication of complex shapes, it still poses a few constraints, e.g., regarding feature size, enclosed voids, and overhang surfaces. Ideally such constraints shall be incorporated into the optimization process. Otherwise a post-process might become necessary, and this process can counteract optimality. In our work, we have taken into account the minimum feature size by the well-established projection method~\cite{Guest04,Wang10}. Concerning the enclosed voids which might trap unsintered powder in SLS (Selective Laser Sintering), in Section~\ref{subsec:truss-vs-wall} we have analyzed parameters which influence the formation of wall-like structures. Nevertheless, a rigorous formulation to guarantee enclosed-void-free is currently out of reach. Regarding overhang avoidance, impressive progress has been made recently by embedding corresponding constraints into the density-based topology optimization~\cite{Langelaar16,Gaynor16,Qian16}. These methods are compatible to our formulation, yet we leave the integration as future work.}

\begin{figure}[t]
\centering
\includegraphics[width=0.98\linewidth]{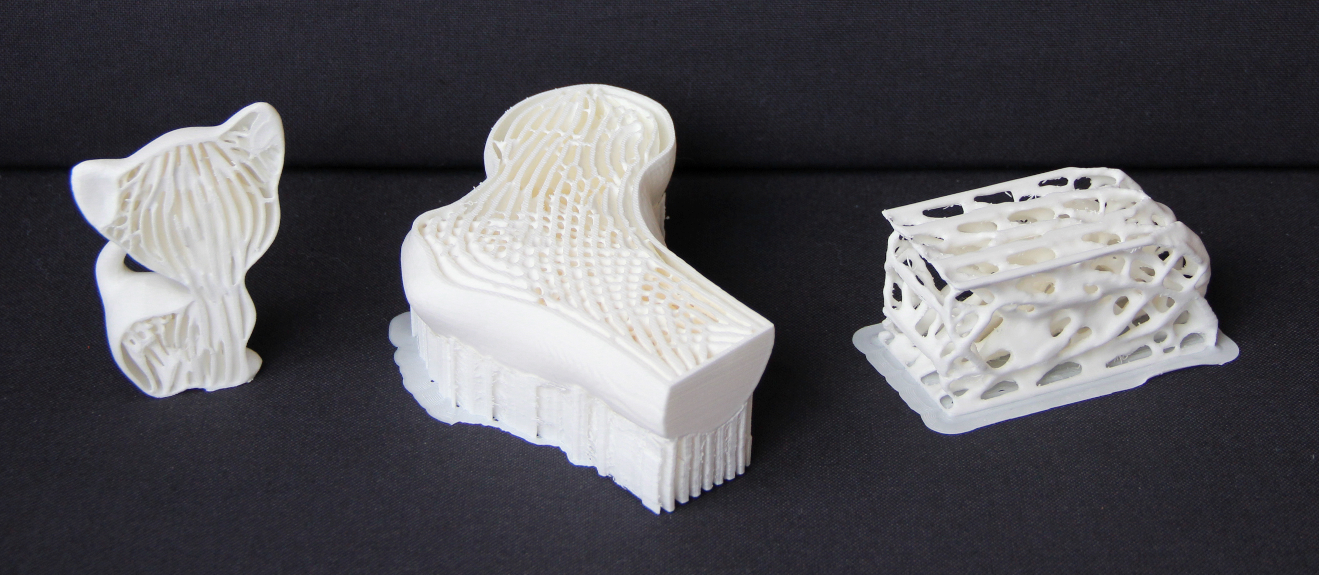}
\caption{
{\revise{}{FDM printed replicas of some models generated by our formulation.}}
}\label{fig:prints}
\end{figure}

\revise{}{To verify the manufacturability of bone-like infills, we have fabricated the femur model using the SLS process (see Fig.~\ref{fig:femur}) and three additional  models using more affordable FDM (Fused Deposition Modelling) printers (see Fig.~\ref{fig:prints}). When using FDM printers, the models were printed without supports for the infill, since, in general, the bone-like infills show small overhang areas that are within the tolerance in FDM printing. The absence of supports leaves a few visual artefacts and unguaranteed mechanical property.}

\section{Conclusion}
\label{sec:conclusion}

We have presented a structural optimization method for obtaining stiffness optimized porous structures. These numerically optimized structures visually resemble trabecular bone, which is lightweight and robust with respect to material deficiency and force variations. This makes the optimized interior structures an ideal candidate for application-specific infill in additive manufacturing. 

\appendix
\section{Appendix}
\label{sec:appendix}
In numerical optimization, the gradient of the objective $c$ and the constraint $g$ with respective to the design variable $\phi$ is needed. It is calculated using the chain rule as follows
\begin{align}
\frac{\partial c}{\partial \phi_e} & = \sum\limits_{i \in \mathbb{M}_e} \left( \frac{\partial c}{\partial \rho_i} \frac{\partial \rho_i}{\partial \tilde{\phi_i}} \frac{\partial \tilde{\phi_i}}{\partial \phi_e} \right), \\
\frac{\partial g}{\partial \phi_e} & = \sum\limits_{i \in \mathbb{M}_e} \left( \sum\limits_{j \in \mathbb{N}_i} \left( \frac{\partial g}{\partial \overline{\rho}_j} \frac{\partial \overline{\rho}_j}{\partial \rho_i} \right) \frac{\partial \rho_i}{\partial \tilde{\phi_i}} \frac{\partial \tilde{\phi_i}}{\partial \phi_e} \right).
\end{align}
The derivative $\frac{\partial c}{\partial \rho_i}$ is calculated using the adjoint analysis
\begin{equation}
\frac{\partial c}{\partial \rho_i} = -\gamma \rho^{\gamma-1}_i (E_0 - E_{\min})u^{\mathrm{T}}_i k_0 u_i.
\end{equation}
The other components can be derived as
\begin{align}
\frac{\partial \rho_i}{\partial \tilde{\phi_i}} &= \frac{\beta(1.0-\tanh^2(\beta(\tilde{\phi}_i-\frac{1}{2})))}{2 \tanh(\frac{\beta}{2})}, \\
\frac{\partial \tilde{\phi_i}}{\partial \phi_e} &= \frac{\omega_{e,i}}{\sum \limits_{k \in \mathbb{M}_i} \omega_{k,i}}, \\
\frac{\partial g}{\partial \overline{\rho}_j} &= \frac{1}{\alpha n} { \left( \frac{1}{n} \sum_{e} \overline{\rho}^p_e \right) ^{\frac{1}{p} - 1} } {\overline{\rho}_j}^{p-1}, \\
\frac{\partial \overline{\rho}_j}{\partial \rho_i} &= \frac{1}{\sum \limits_{k\in \mathbb{N}_j} 1}. 
\end{align}

\section*{Acknowledgements}
The authors gratefully acknowledge the support from the H.C. {\O}rsted Postdoc Programme at the Technical University of Denmark, which has received funding from the People Programme (Marie Curie Actions) of the European Union's Seventh Framework Programme (FP7/2007-2013) under REA grant agreement no. 609405 (COFUNDPostdocDTU), and the support from the Villum foundation through the NextTop project. 

\bibliographystyle{IEEEtran_noURL}
\bibliography{topopt}

\vspace{-0.4in}
\begin{IEEEbiography}[{\includegraphics[width=1in,height=1.25in,clip,keepaspectratio]{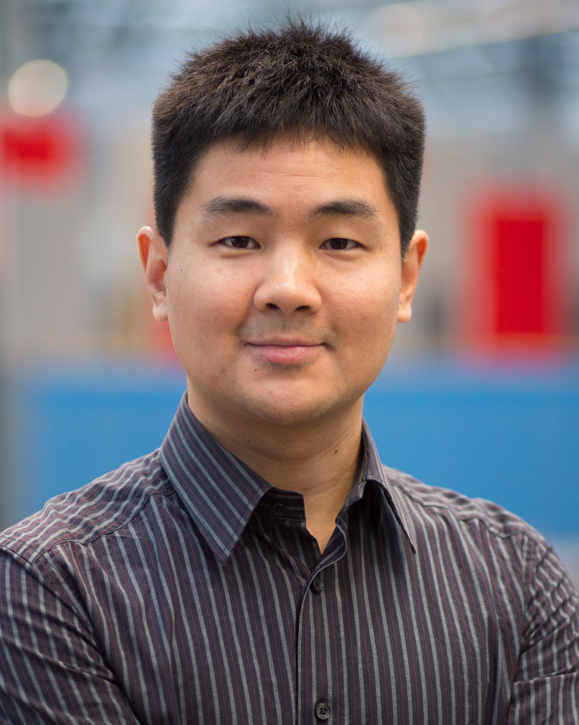}}]{Jun Wu}
is a PostDoc at the Department of Mechanical Engineering at the Technical University of Denmark. He received a PhD in Computer Science in 2015 from the Technical University of Munich, Germany, and a PhD in Mechanical Engineering in 2012 from Beihang University, Beijing, China, where he also received a B.Eng in Astronautics Engineering in 2006. His research is focused on geometric and physical modeling, with applications in surgical simulation and design optimization.
\end{IEEEbiography}
\vspace{-0.4in}
\begin{IEEEbiography}[{\includegraphics[width=1in,height=1.25in,clip,keepaspectratio]{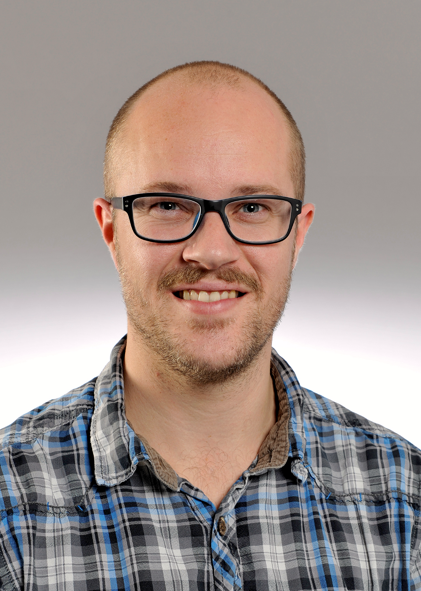}}]{Niels Aage} is an Associate Professor at the 
Department of Mechanical Engineering at 
the Technical University of Denmark. He received his 
PhD in Optimal Design in 2011 from the Technical 
University of Denmark. His research interest is focused on the area of 
large scale, parallel numerical methods for PDE constrained
optimization, with emphasis on topology and shape optimization
problems in multiphysical settings.
\end{IEEEbiography}
\vspace{-0.4in}
\begin{IEEEbiography}[{\includegraphics[width=1in,height=1.25in,clip,keepaspectratio]{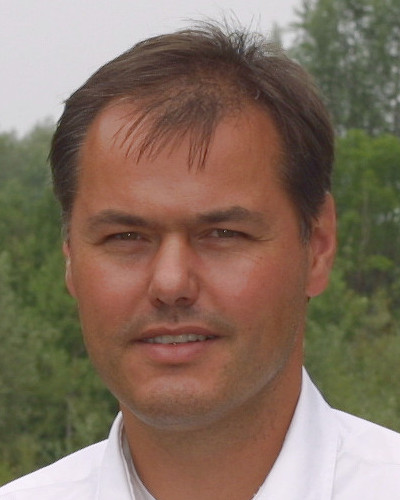}}]{R\"udiger Westermann}
studied computer science at the Technical University Darmstadt, Germany. He pursued his Doctoral thesis on multiresolution techniques in volume rendering, and he received a PhD in computer science from the University of Dortmund, Germany. In 2002, he was appointed the chair of computer graphics and visualization at the Technical University Munich. His research interests include scalable simulation and visualization algorithms, GPU computing, real-time rendering of large data, and uncertainty visualization.
\end{IEEEbiography}
\vspace{-0.4in}
\begin{IEEEbiography}[{\includegraphics[width=1in,height=1.25in,clip,keepaspectratio]{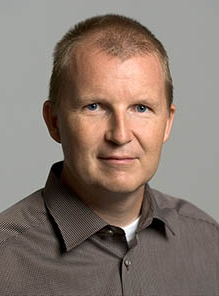}}]{Ole Sigmund} is a Professor at the Department of Mechanical Engineering, Technical University of Denmark (DTU). He obtained his Ph.D.-degree 1994 and Habilitation in 2001 and has had research positions at University of Essen and Princeton University. He is a member of the Danish Academy of Technical Sciences and the Royal Academy of Science and Letters (Denmark) and is the former elected President (2011-15, now EC member) of ISSMO (International Society of Structural and Multidisciplinary Optimization). Research interests include theoretical extensions and applications of topology optimization methods to mechanics and multiphysics problems.
\end{IEEEbiography}

\end{document}